\documentclass[a4paper]{article}

\usepackage{authblk}
\usepackage[utf8]{inputenc}
\usepackage[T1]{fontenc}
\usepackage{pslatex}
\usepackage[english]{babel}
\usepackage{fancyhdr}
\usepackage{hyperref}
\usepackage{amsmath,amssymb,amsfonts}
\usepackage{graphicx}
\usepackage{amsmath}
\usepackage{amsthm}
\usepackage{amsfonts}
\usepackage{amssymb}
\usepackage{algorithm}
\usepackage[noEnd=false]{algpseudocodex}
\usepackage{microtype}
\usepackage{rotating}
\usepackage{todonotes}
\usepackage{tikz}
\usepackage{xparse}
\usepackage{appendix}
\usepackage{url}
\usepackage{tabularx}
\usepackage{paralist}
\usepackage{booktabs}
\usepackage{multirow}
\usepackage{subcaption}
\usepackage{cancel}
\usetikzlibrary{calc,shapes.callouts,shapes.arrows}
\usetikzlibrary{automata}
\usetikzlibrary{positioning}
\usetikzlibrary{decorations.pathreplacing}
\usetikzlibrary{decorations.pathmorphing}
\usetikzlibrary{arrows}
\usetikzlibrary{shapes}

\newtheorem{definition}{Definition}

\newtheorem*{problem*}{Problem}

\providecommand{\keywords}[1]{\textbf{\textit{Keywords }} #1}

\begin{document}

\title{Smooth Number Message Authentication Code in the IoT Landscape}

\author[1]{Eduard-Matei Constantinescu}
\author[1]{Mohammed Elhajj}
\author[1]{Luca Mariot}

\affil[1]{{\normalsize Semantics, Cybersecurity and Services Group, University of Twente, 7522 NB Enschede, The Netherlands} \\
	
	{\small \texttt{e.constantinescu@student.utwente.nl, m.elhajj@utwente.nl, l.mariot@utwente.nl}}}
	
\maketitle

\begin{abstract}
This paper presents the Smooth Number Message Authentication Code (SNMAC) for the context of lightweight IoT devices. The proposal is based on the use of smooth numbers in the field of cryptography, and investigates how one can use them to improve the security and performance of various algorithms or security constructs. The literature findings suggest that current IoT solutions are viable and promising, yet they should explore the potential usage of smooth numbers. The methodology involves several processes, including the design, implementation, and results evaluation. After introducing the algorithm, provides a detailed account of the experimental performance analysis of the SNMAC solution, showcasing its efficiency in real-world scenarios. Furthermore, the paper also explores the security aspects of the proposed SNMAC algorithm, offering valuable insights into its robustness and applicability for ensuring secure communication within IoT environments.
\end{abstract}


\section{Introduction}
\label{introduction}
In the expansive domain of the Internet-of-Things (IoT), which encompasses a vast array of devices ranging from minuscule sensors to substantial mechanical units, securing communication is paramount to preserve the confidentiality, integrity, and availability of data \cite{vailshery_2016}. This urgency is underscored by challenges unique to the IoT landscape, such as the absence of universal security standards and regulations, substantial disparities in hardware and software configurations, and notably, the constraints imposed by limited computational power and resources \cite{DBLP:journals/fi/ElhajjMF23}. To authenticate communication between these diverse IoT devices, two distinct approaches have emerged: a symmetric method employing HMAC (Hash-based Message Authentication Code) and an asymmetric one utilizing certificates and signatures \cite{DBLP:conf/giots/Suarez-AlbelaFF18}. Studies have explored various optimizations within these methods to enhance their efficiency. For instance, researchers have proposed an ECC-based OTP authentication scheme specifically tailored for IoT, demonstrating reduced power consumption and energy usage \cite{DBLP:journals/sj/HammiFKZB20}. Conversely, within the symmetric realm, efforts have led to the development of energy-efficient HMAC variants, such as a lightweight HMAC authentication solution and innovative approaches like Secure Vaults and "energy-efficient" HMAC, each designed to optimize energy consumption significantly \cite{DBLP:conf/green/CastellonRKDB22}. These advancements underscore the ongoing efforts to balance the security needs of the IoT ecosystem with the inherent limitations of its diverse devices, fostering innovations that pave the way for secure, energy-efficient communication within this dynamic landscape.

In this paper, we follow a different perspective by proposing the Smooth Numbers Message Authentication Code (SNMAC), a variation of Message Authentication Codes (MAC) tailored for IoT devices. Unlike HMAC, SNMAC is based on the use of \textit{smooth numbers} \cite{granville2008smooth} to enhance performance in resource-constrained IoT environments.
In number theory, smooth numbers are defined as positive integers devoid of prime factors greater than a specific value. These numbers find applications in various mathematical algorithms, including the Fast Fourier Transform \cite{cooley1965algorithm} and integer factorization \cite{briggs1998introduction}. In cryptography, smooth numbers are interesting both for the design of secure cryptographic schemes and their cryptanalysis. Hash functions based on "very smooth" numbers have been designed and proven to be collision resistant, making them particularly appealing for safeguarding the integrity of messages~\cite{DBLP:conf/eurocrypt/ContiniLS06}.

\subsubsection*{Research questions} \label{research_questions}

Given that the IoT landscape struggles with hardware limitation issues and, at the same, requires secure communication channels when exchanging messages, this study aims to explore the following research questions:

\begin{enumerate}

    \item[\textbf{RQ1}] \textit{How can smooth numbers be used to enhance the security and performance of a Message Authentication Code algorithm for lightweight IoT devices?}
    \item[\textbf{RQ2}] \textit{What is the performance of a Smooth Number Message Authentication Code for lightweight IoT devices in comparison with existing MACs constructions?}
\end{enumerate}

\subsubsection*{Contributions} \label{objectives}
While the idea of using smooth numbers for optimisation purposes in the IoT environment is novel, the literature can provide valuable knowledge about two crucial adjacent topics: the requirements of lightweight authentication solutions for IoT devices and the potential usage of smooth numbers in optimising efficiency and performance. Thus, the main contributions of this paper are summarized as follows:

\begin{itemize}
    \item We describe in detail the building blocks and the design of the SNMAC algorithm.
    \item We investigate the feasibility of SNMAC for IoT devices, leveraging the existing knowledge on smooth numbers.
    \item We analyse a prototype solution of the SNMAC algorithm in terms of performance and security.
\end{itemize}

The rest of this paper is structured as follows. Section \ref{related_work} presents an analysis of the relevant literature. Section \ref{proposed_solution} describes the design of the SNMAC algorithm. Section \ref{performance_analysis} concerns the performance analysis of the SNMAC, while Section \ref{security_analysis} concerns the security analysis. Finally, Section \ref{conclusion} shortly presents the findings of the research and points out some directions for future improvements.

\section{Related Work}
\label{related_work}
This section focuses on the research encompassing smooth numbers within the realm of cryptography, exploring their potential to enhance or replace current cryptographic methods.

In the study introduced by in~\cite{DBLP:journals/tc/DimitrovVA22}, the authors leveraged smooth integers to expedite RSA key generation. Traditionally, RSA key generation involves a two-step process: initially, trial divisions are performed to filter out integers with small prime factors (typically up to the first 100 primes), followed by primality tests like the Miller-Rabin test to further ensure primality. The authors identified a gap in the existing methodology: the initial step could be significantly enhanced by considering a larger set of prime factors (up to the first $10,000$ primes), thereby increasing the likelihood of generating a large prime number. However, this approach involved computationally expensive trial divisions, which can hinder the generation process. To overcome this, the authors proposed using smooth numbers as an alternative. They introduced a one-way function based on the problem of reversing the sum of two smooth integers, exploiting the fact that the sum of two $n$-smooth integers remains indivisible by any primes up to the $n^{th}$ prime. By efficiently generating large smooth integers, the authors achieved notable results, reducing execution time by up to 12\% during the large prime generation process. 

In the study conducted by the authors of \cite{aragona2016real}, an assessment was made on RSA keys used by Certificate Authorities (CAs) in financial transactions. They employed the \textit{General Number Field Sieve} algorithm to factor large RSA moduli by identifying smooth numbers, which could potentially be factors. The authors developed an efficient approach to systematically factor the RSA modulus and proposed a formula to evaluate the feasibility of such an attack, taking into account factors such as the attacker's budget, RSA modulus size, and available time. The study concluded that factoring a 1024-bit RSA modulus within a reasonable timeframe would require an exponentially increasing budget concerning the available time.

Certainly, building upon the previous studies, the authors of \cite{karthi2019enhanced} introduced a modified version of the Very Smooth Hash called the Very Smooth Discrete Logarithm hash (VSDL). Unlike its predecessor, VSDL relies on the classical discrete logarithm problem rather than the nontrivial modular square root problem. This variant operates as an \emph{S-bit hash function}, with its security strength linked to the complexity of finding discrete logarithms of primitive roots of smooth numbers modulo an S-bit composite. The authors conducted a series of experiments to assess the performance and security of the algorithm. Their findings indicated that VSDL surpassed other hashing methods in terms of time efficiency and demonstrated superior security performance, particularly in scenarios involving collision and inversion attacks, when compared to similar algorithms.

The research detailed in \cite{DBLP:conf/nspw/Durmuth13} presented a new method for password hasing, named \textit{Useful Password Hashes} (UPHs). The primary objective of UPHs was to optimize computing cycles by simultaneously addressing additional computational problems during the hashing process, thereby minimizing wasted resources. The study proposed three distinct UPH constructions, each rooted in diverse cryptographic challenges: brute-forcing block-ciphers, discrete logarithms, and integer factorization. Notably, in the latter construction, the security of the approach relied on solving the \textit{quadratic sieve} problem, specifically finding $a \neq b$ such that $a^2 = b^2 \mod N$. This problem was translated into locating solutions to a specific polynomial $Q(x)$, where $x$ represents a smooth number. This  approach enabled the construction to perform smooth number "sieving" operations during the generation and evaluation of password hashes. While the paper provided a rudimentary security analysis, ensuring parameter safety and algorithm correctness, it omitted considerations of common attacks against hashing methods. Furthermore, no experimental evidence was presented to substantiate the performance claims made.

In the research discussed in \cite{DBLP:journals/ijnsec/PadmavathyB15}, novel methods were explored to solve the discrete logarithm problem using smooth numbers. Specifically, the study focused on addressing this problem for safe primes, which are primes of the form $p=2q + 1$. The researchers extended the concept of smooth numbers to finite fields, introducing \textit{smooth factor bases}. They developed mathematical relationships among smooth numbers, smooth factor bases, and soft primes, enabling the computation of discrete logarithm solutions within specific ranges. Experimental results showcased the successful application of these techniques, demonstrating their effectiveness in solving discrete logarithm problems for various types and sizes of primes and group orders.

In the study conducted by \cite{DBLP:conf/tgc/AmarilliHBMNR12}, the focus was on the problem of set reconciliation, which involves synchronizing fixed-sized value multisets with minimal transmission complexity. This problem can be viewed as a translation of the file synchronization challenge, which revolves around validating the file hierarchy of a remote host in the presence of outdated file versions. The researchers introduced a novel number-theoretic reconciliation protocol called \textit{Divide and Factor} (D\&F), aiming for optimal asymptotic transmission complexity under ideal circumstances. The D\&F protocol, detailed in the study, was complemented by hashing techniques for file reconciliation tasks. Notably, the authors explored the potential of generating smooth numbers instead of large primes, highlighting the efficiency gains. Experimental results were presented, demonstrating that the proposed algorithm excelled in efficient transmission, albeit with increased computational workload.

\section{Proposed Solution} \label{proposed_solution}

We now introduce our proposed solution for a smooth-number based MAC algorithm, based on the literature review results presented earlier. First, we identify what the SNMAC is designed to achieve in the IoT environment. Then, we shortly discuss relevant aspects of the HMAC construction and other relevant details learned from the IoT and smooth numbers related research papers. Finally, the design of the SNMAC algorithm is presented.

As discussed in previous sections, based on literature findings, most MAC algorithms designed for IoT devices try to improve classical constructs by introducing more efficient computation tasks or exploring niche alternatives to existing MAC components. Furthermore, the literature review also shows that smooth numbers can be used to build efficient constructs when dealing with large numbers, be it factorization or generation.

\subsection{Design Decisions}

The proposed solution strives to achieve the efficiency goals while providing precise implementation details and proper discussions. As shown in the literature, smooth numbers can increase the efficiency and performance of various algorithms or security constructs \cite{DBLP:journals/ijnsec/PadmavathyB15,DBLP:conf/tgc/AmarilliHBMNR12,aragona2016real,DBLP:journals/tc/DimitrovVA22,DBLP:conf/nspw/Durmuth13}. The proposed SNMAC incorporates such an efficient, smooth number based construction in its design. Further, the SNMAC algorithm uses the existing design rules of HMAC algorithms as initial building blocks. This decision has various reasons: 

\begin{itemize}
    \item HMAC is a symmetric construction, making it faster than any public-key alternatives.
    \item HMAC uses a modest yet robust hashing construction, making it easy to adapt or change.
    \item HMAC is easy to implement on various hardware and software architectures.
\end{itemize}

\subsection{Key Generation}

Since SNMAC is inspired by the HMAC design, it inherits HMAC's secret key generation and usage properties. First, the secret key must be efficiently generated/chosen from an ample key space and then shared between the sender and the receiver. In the SNMAC design, we assume that the sender manages the secret key generation procedure and explores how the sender can handle this task efficiently. However, the SNMAC design does not concern how the secret key is exchanged between the communicating parties. This task can be achieved using algorithms such as the Diffie-Hellman key exchange protocol. 

In a typical HMAC protocol, the secret key has no crucial requirements other than choosing a proper key length (usually in the 128-256 bit range). However, the SNMAC algorithm will use secret keys in the form of RSA moduli since its hashing function requires the unique properties of such moduli. This requirement introduces a new challenge: how can one efficiently generate an RSA modulus of varying bit length? 

First, we recall the definition of the RSA modulus: $n$ is an RSA modulus if $n = pq$, where $p$ and $q$ are two large prime numbers. Typically, $p$ and $q$ are randomly chosen, and the standard method for choosing them is to randomly generate large integers and use a primality test until two large primes are found. However, Dimitrov et al. \cite{DBLP:journals/tc/DimitrovVA22} discuss how this approach is inefficient and how most academic endeavours focus on the primality tests, disregarding possible improvements to the random integer generation procedure. To address this issue, the authors provide an elegant algorithm for generating large random numbers not divisible by a set of small primes up to a given limit.

The logic behind it is quite elementary. Let us assume that we wish to generate a random number that is not divisible by the first 50 primes; in other words, the set of primes $S = [2, 3, 5, ..., 223, 227, 229]$. We start by dividing the set into two smaller disjoint subsets. One way of doing so is by putting every other prime from $S$ in $S_1$ and the remaining primes in $S_2$. Thus, $S_1 = [2, 5, 11, ..., 211, 227]$ and $S_2 = [3, 7, 13, ..., 223, 229]$. Then, we generate two smooth numbers $a$ and $b$:
\[ a = 2^{r_1} \cdot 5^{r_3} \cdot 11^{r_5} ... \]
\[ b = 3^{r_2} \cdot 7^{r_4} \cdot 13^{r_6} ... \]

According to the authors, the exponents $r_i$ are chosen from a set of random integers in the $[1,4]$ interval, which provides sufficient randomness for this integer generation procedure.

Now, it is evident that the sum $x = a + b$ is not divisible by any of the first 50 primes in $S$. Thus, $x$ has a higher chance of being a random prime when compared to a randomly selected integer of a similar bit size. To ensure the validity of this claim, we continue by executing a primality test, such as Miller-Rabin, to state with a high degree of confidence whether or not $x$ is a prime. Furthermore, the authors provide an experimentally found boundary for the number of small primes required to generate a random large number: to generate a $k$-bit number, it is sufficient to use the first $k / \ln{k} $ primes. Thus, a 512-bit prime can be generated using only the first 82 primes, and a 2048-bit prime using only the first 269 primes. We now have all the required components to properly generate a secret key; the pseudocode of our SNMAC key generation procedure is reported in Algorithm~\ref{alg:keygen}.
\begin{algorithm}[t]
    \caption{SNMAC secret key generation} \label{alg:secret_key}
    \label{alg:keygen}
    \begin{algorithmic}
        \Require $k > 2$
        \Ensure $p \neq q$
        \Ensure $n = p \cdot q$ is a $2k$-bit RSA modulus

        \Function{{\sc Generate-Secret-Key}}{$k$}   
            \State $p \gets$ {\sc Generate-Prime}$(k)$
            \State $q \gets p$
            \While{$q=p$}
                \State $q \gets$ {\sc Generate-Prime}$(k)$
            \EndWhile
    
            \State $n \gets p \cdot q$ \Comment{$n$ is an RSA modulus}
            
            \State \Return n 
        \EndFunction
    \end{algorithmic}
    
\end{algorithm}
The structure of the pseudocode is quite straightforward: we rely on the algorithm introduced in~\cite{DBLP:journals/tc/DimitrovVA22} as a subroutine {\sc Generate-Prime} to sample the first candidate prime number $p$ of the RSA modulus. Then, a while loop is used to call iteratively {\sc Generate-Prime} on the same parameter, until the second value of $q$ of the RSA modulus is different from $p$.

\subsection{Hashing Function}
\label{hashing_function}
HMAC~\cite{DBLP:journals/rfc/rfc2104} is a tool for message authentication that uses an arbitrary cryptographic hash function, such as  SHA-2 or SHA-3. As mentioned before, SNMAC builds upon the design principles of HMAC and aims to use a different hashing function that relies on specific properties of smooth numbers. First introduced in 2005, VSH is a cryptographic hash function constructed on the \textit{Very Smooth Number Nontrivial Modular Square Root} problem~\cite{DBLP:conf/eurocrypt/ContiniLS06}. We introduce the definitions related to Very Smooth Numbers and Quadratic residues below. 

\begin{definition}[Very Smooth Number (VSN)] \label{def:vsn}
    The integer $m$ is a \textbf{Very Smooth Number} if, for a fixed constant $c$ and an integer $n$, the largest prime factor of $m$ is $(\log_n)^c$.
\end{definition}

\begin{definition}[Very Smooth Quadratic Residue] \label{def:vsqr}
    The integer $b$ is a \textbf{Very Smooth Quadratic Residue} modulo $n$ if $b$ is a Very Smooth Number and there exists an integer $x$ such that $b \equiv x^2 \mod n$.
\end{definition}

Since trivial square roots are easy to compute using searching algorithms in particular fields, such as the real field, they are unsuitable for our cryptographic construction. Hence, we are only interested in non-trivial square roots, i.e. those where $x^2 \geq n$. 

We can now define the VSN Nontrivial Modular Square Root Problem, also shortened as VSSR:

\begin{definition}[Very Smooth Number Nontrivial Modular Square Root Problem (VSSR)] \label{def:vssr}
    Let $n=pq$ be an RSA modulus and let $k \leq (\log_n)^c$. Given $n$, find $x \in \mathbb{Z}_{n}^{*}$ such that $x^2 \equiv \prod_{i=1}^{k} p_i^{e_i}$, where $p_i$ is prime for each $i \in  \{1,\ldots, k\}$, and at least one of $e_1, \ldots, e_k$ is odd.
\end{definition}

The difficulty of the VSSR problem comes from the assumption that solving it is as complex as factoring a hard-to-factor $l$-bit modulus, where $l$ is roughly smaller than the size of $n$. Starting from Definition~\ref{def:vssr}, the authors of~\cite{DBLP:conf/eurocrypt/ContiniLS06} designed the Very Smooth Hash compression function, whose pseudocode is reported in Algorithm~\ref{alg:vsh}.

\begin{algorithm}[t]
    \caption{Very Smooth Hash (VSH) \cite{DBLP:conf/eurocrypt/ContiniLS06}} 
    \label{alg:vsh}
    \begin{algorithmic}
        \Require $n$ is large RSA modulus 
        \Require $m \gets [m_1, m2_, ..., m_l]$ \Comment{$l$-bit message to be encoded}
        \Require $k$ largest integer such that $\prod_{i=1}^k p_i < n$
        \Require $l < 2^k$

        \Function{VSH}{n, m}
            \State $x_0 \gets 1$
            \State $L \gets$ smallest integer $\geq l/k$ \Comment{number of blocks}
            \State \verb|PAD| $\gets [0, 0, \dots, 0]$ such that $\verb|PAD| = Lk - l$ 
            \State $m \gets m + \verb|PAD|$
            \State $z \gets$ binary representation of $l$
            \State $m \gets m + z$
    
            \For{j=0, \dots, L}
                \State $x_{j+1} = x_j^2 \prod_{i=1}^k p_i^{m_{jk + i}} \mod n$ \Comment{compression function}
            \EndFor
    
            \Return $x_{L+1}$ \Comment{the tag associated with message $m$}
        \EndFunction
    \end{algorithmic}
\end{algorithm}

Before we move on to the actual implementation of the SNMAC, it is essential to highlight some important aspects of the VSH pointed out by the authors of~\cite{DBLP:conf/eurocrypt/ContiniLS06}:

\begin{itemize}
    \item Unlike other hashing functions, the VSH has no fixed tag length. Instead, the tag length is equal to the size of the secret RSA modulus (in bits).
    \item The VSH scheme has a strong requirement for the secret key to be an VSH modulus and will only function properly if the requirement is met.
    \item The VSH is collision-resistant based on the VSSR assumption.
    \item VSH satisfies a multiplicative property.
    \item Asymptotically, VSH requires a single multiplication per $\log(n)$ message bits, making it useful in embedded devices where coding space is restricted.
\end{itemize}

\subsection{Overall MAC Algorithm} \label{blueprint}
After defining how to generate secret keys and how to hash messages, we can describe the general design of our SNMAC algorithm. The pseudocode is reported in Algorithm~\ref{alg:snmac}.

\begin{algorithm}[t]
    \caption{SNMAC} 
    \label{alg:snmac}
    \begin{algorithmic}
        \Function{SNMAC}{$n, m$}
            \State $l \gets |n|$ \Comment{the bit-length of $n$}
            \State \verb|OPAD| $\gets 01011100 \dots$ such that |\verb|OPAD|| $ = l$
            \State \verb|IPAD| $\gets 00110110 \dots$ such that |\verb|IPAD|| $ = l$
            \State \verb|TAG| $\gets \verb|VSH|((n \oplus \verb|OPAD|) || \verb|VSH|((n \oplus \verb|IPAD|) || m))$

            \Return \verb|TAG|
        \EndFunction
    \end{algorithmic}
\end{algorithm}
As it can be seen, SNMAC follows the same structure of HMAC, with the main difference that VSH is the underlying compression function. Hence, the tag of a message $m$ is computed first evaluating VSH over the bitwise XOR of an inner pad and the modulus $n$. The result is then concatenated with the bitwise XOR of the modulus and the outer pad, and with the message. Finally, VSH is computed over this concatenation.

\begin{figure}[t]
    \centering
    \includegraphics[width=0.7\textwidth]{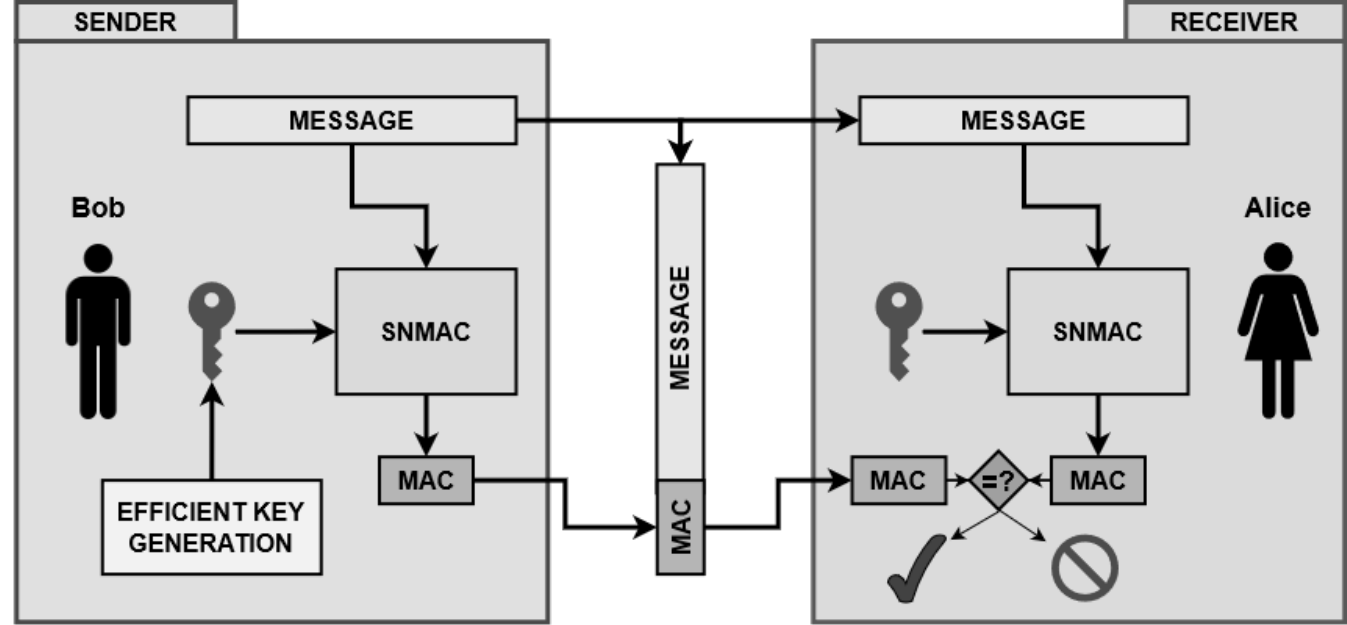}
    \caption{SNMAC scheme}
    \label{fig:snmac}
\end{figure}

Figure \ref{fig:snmac} showcases how SNMAC can be used to authenticated messages sent by Bob (the sender) and Alice (the receiver). First, Bob generates a secret key of length $l$ and shares it with Alice. As mentioned before, the key exchange method could be any standard public-key protocol, such Diffie-Hellman. After both parties agree on the secret key, Bob generates a tag for his message $m$ using the SNMAC algorithm. Afterwards, Bob sends Alice his message together with its associated tag. After receiving them, Alice computes a tag of her own, using the received message and the secret key agreed upon previously. If Alice's tag matches Bob's, the check is successful.

\section{Performance Analysis} 
\label{performance_analysis}
The performance analysis of the authentication algorithm encompasses an evaluation of various key factors that directly influence its efficiency. The algorithm's CPU usage provides insights into how efficiently it utilises processing power during key generation and message hashing operations. Memory usage is another important aspect considered in the performance analysis, and it involves assessing the algorithm's system memory utilisation for storing and manipulating data during its lifetime. Power consumption is crucial, especially in resource-constrained IoT environments or IoT devices with limited battery life. Analysing the algorithm's power consumption helps estimate its impact on energy efficiency and the overall system performance.

\subsection{Experimental setup}
\label{setup}
The performance of the proposed SNMAC algorithm was evaluated through a comprehensive series of experiments that involved benchmarking a prototyped implementation. An experimental setup was devised to ensure accurate and reliable results, encompassing both hardware and software components. Recognizing that the SNMAC algorithm is specifically designed for resource-constrained environments, creating an experimental environment that closely replicated such conditions was necessary. This approach provided a more realistic assessment of the algorithm's performance under the intended operational constraints. 

In terms of hardware, a careful selection was done to emulate resource limitations. This involved using hardware components with comparable specifications to those commonly found in resource-constrained systems. For example, processors with limited computing power, constrained memory resources, and lower power budgets were employed. Additionally, the software configuration played a significant role in simulating a resource-constrained environment. The prototyped implementation of the SNMAC algorithm was run on an operating system configured to mimic the conditions typically encountered in such environments.

The experiments conducted within this setup included various scenarios and workloads to thoroughly assess the performance of the algorithm. We took measurements to analyze CPU usage, memory consumption, power consumption, and other relevant performance metrics. By subjecting SNMAC to rigorous testing in a simulated resource-constrained environment, our evaluation aimed to provide valuable insights into the behaviour of the algorithm under real-world conditions. The benchmarking process involved comparing the performance of the SNMAC algorithm against established authentication algorithms widely used in similar environments. This comparative analysis helped ascertain the strengths and weaknesses of the proposed algorithm, shedding light on its potential advantages and trade-offs in resource-constrained scenarios. 

\subsubsection{Hardware and Software. }

We implemented the prototype on a standalone Raspberry Pi 3 device, chosen for both its popularity and due to its relatively limited available resources compared to regular computing devices. The Raspberry Pi 3 ad no modifications and performed in an isolated environment. The second hardware device involved in the experimental setup is a USB power monitor tool that collects power consumption data during the Raspberry Pi's running time. This power monitor device can detect and record various relevant data, such as values for the device voltage, flow of electric charge, electrical resistance, rate of energy transfer and device temperature.

The software choice includes the operating system and the programming language used to implement and execute the prototype. Since the main hardware of choice is a stock Raspberry Pi 3, the operating system is also the latest version of stock Raspbian OS. Thus, the only relevant choice was the programming language used to successfully implement and benchmark the prototype algorithm. Albeit most IoT and embedded devices tend to favour low-level programming languages, we chose to use Python 3.9 for its flexibility, ease of code implementation and testing, and wide range of publicly available extensions. 

\subsection{Experimental Results}

The performance analysis includes various experiments that recorded relevant metrics regarding the performance of the proposed SNMAC. Where possible, we benchmarked this performance against other popular message authentication codes, most notably the HMAC construction using standard hashing functions such as SHA-2. Furthermore, the experiments tracked the two primary functions of the SNMAC algorithm, namely the key-generation procedure and the hash generation for an input plaintext message. As mentioned in Section \ref{proposed_solution}, the key-sharing protocol is not included in the algorithm's design and, thus, will not be considered in this experimental analysis. Finally, this analysis also does not cover verifying the authenticity of a message-hash pair since this process is almost identical to the message-hash generation process.

\subsubsection{CPU Usage: }

One of the most important metrics for SNMAC performance is CPU usage. It is crucial that, in resource-constrained IoT environments, the proposed algorithm will perform with proper speed and efficiency. 

\begin{figure}[t]
     \centering
         \includegraphics[width=0.8\textwidth]{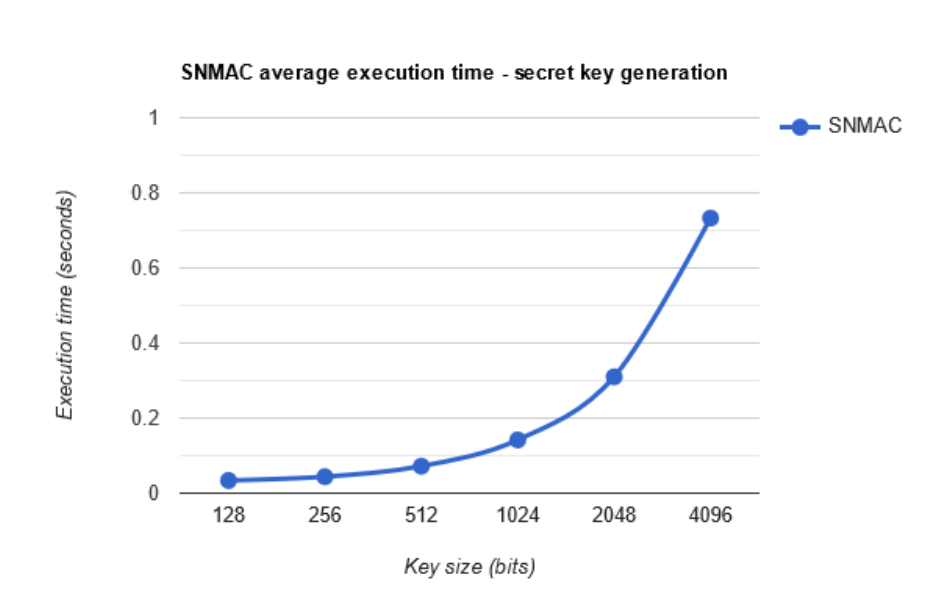}
         \caption{SNMAC average execution time - secret key generation}
         \label{fig:key_generation_time}
\end{figure}

\begin{figure}[t]
         \centering
         \includegraphics[width=0.8\textwidth]{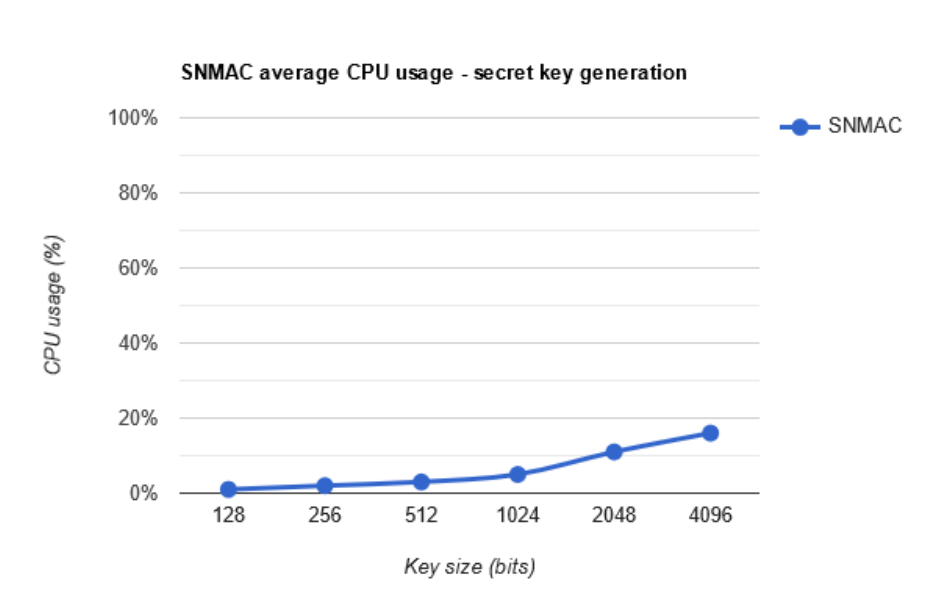}
         \caption{SNMAC average CPU usage - secret key generation}
         \label{fig:key_generation_cpu}
\end{figure}

The first experiment analysed the performance of the key generation procedure for various desired lengths of the secret key, ranging from 128-bit keys up to 4096-bit keys. Each key generation involved efficiently finding two large primes, which together form an RSA modulus. For each key length, the key generation procedure was repeated $10,000$ times. Figure \ref{fig:key_generation_time} plots the average execution time to generate a secret key of a given length, while Figure \ref{fig:key_generation_cpu} plots the average CPU usage for the secret key generation procedure. 

\begin{figure}[t]
         \centering
         \includegraphics[width=0.8\textwidth]{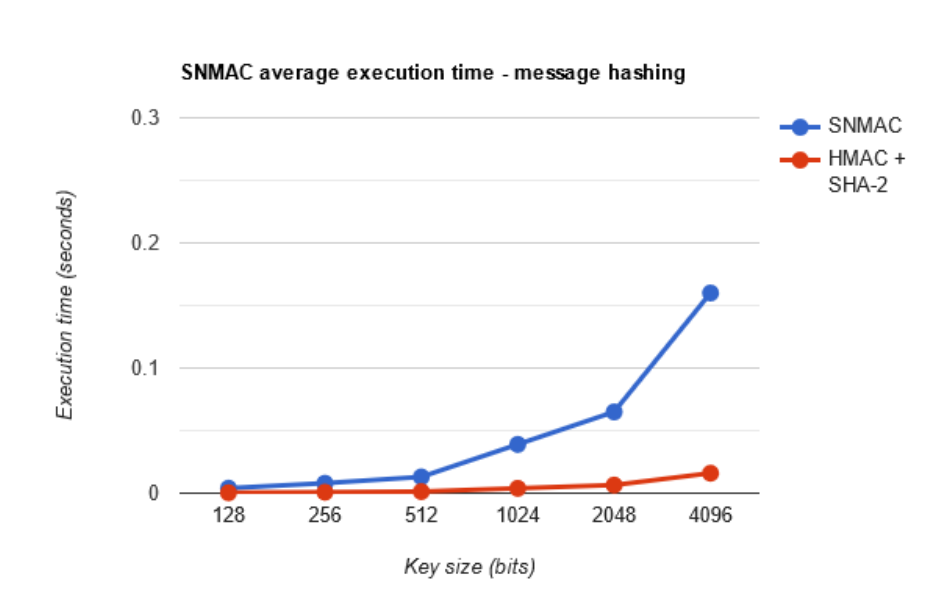}
         \caption{SNMAC average execution time - message hashing}
         \label{fig:hashing_time}
\end{figure}

\begin{figure}[t]
         \centering
         \includegraphics[width=0.8\textwidth]{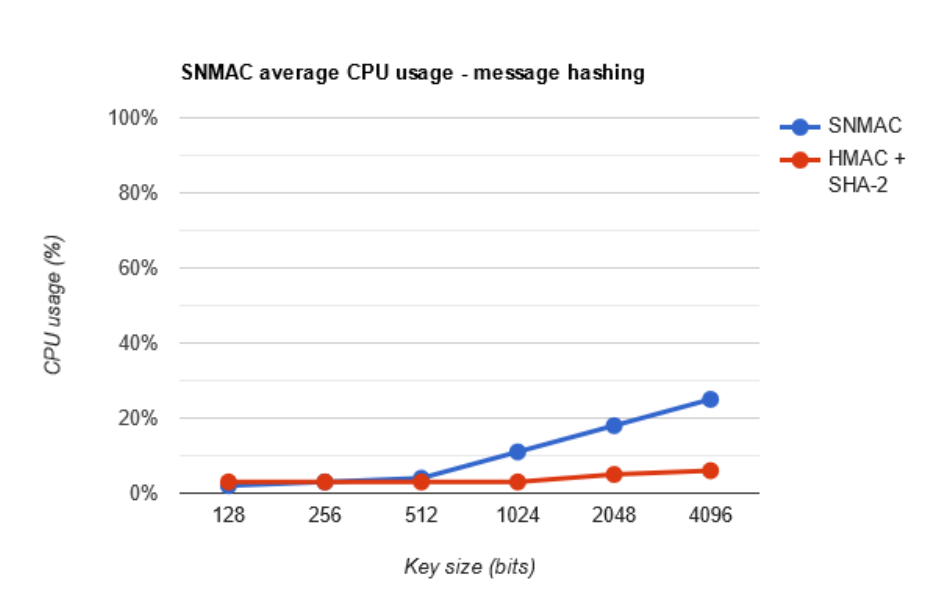}
         \caption{SNMAC average CPU usage - message hashing}
         \label{fig:hashing_cpu}
\end{figure}

The second experiment analysed the performance of generating a message-hash pair for a given plaintext message. Since the SNMAC performance depends on the key length and not the message length, the message was chosen to be a fixed random 128-bit message. Furthermore, the experiment involved generating a hash for the fixed message using precomputed secret keys of various lengths, ranging from 128 up to 4096 bits. The same experiment was repeated using an HMAC construction combined with SHA-2. Figure~\ref{fig:hashing_time} and \ref{fig:hashing_cpu} respectively plot the average execution time and the average CPU usage.

\subsubsection{Memory Usage: }
The next metric to analyse is the memory used when compiling and executing the SNMAC algorithm. Compared to the previous experiment, this analysis observes the algorithm as a whole instead of separating it into the key generation and message hashing procedures. The reasoning behind this decision is that the OS will load the entire algorithm into memory, regardless whether the key generation or message hashing execute. Furthermore, this experiment cannot be benchmarked against an HMAC construction since its implementation was provided by external libraries, meaning that the OS will load into memory more than just the authentication algorithm.

\begin{figure}[t]
    \centering
    \includegraphics[width=1.1\textwidth]{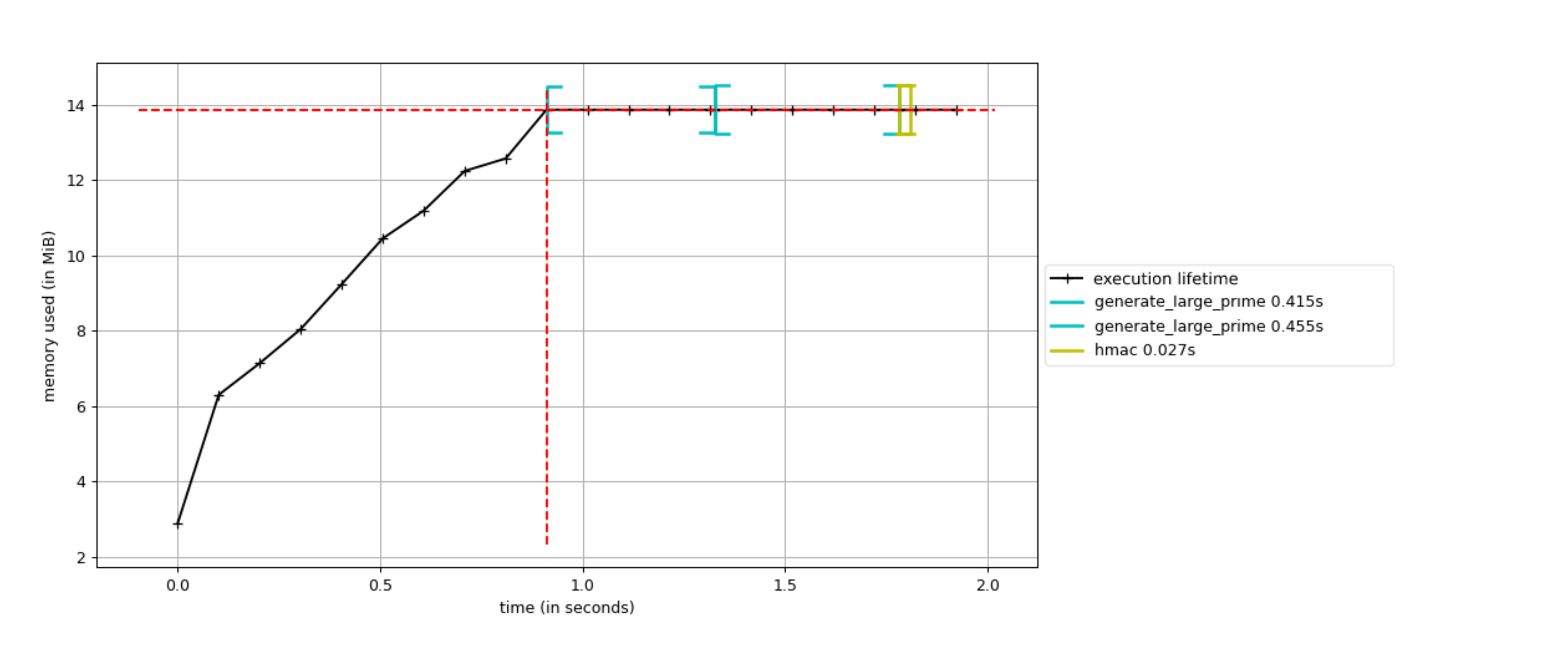}
    \caption{SNMAC memory usage}
    \label{fig:memory_plot}
\end{figure}

Figure \ref{fig:memory_plot} plots the memory used for the SNMAC algorithm during a regular execution, including a secret key generation procedure and a message hashing task. Since experiments involving various message lengths and key lengths returned roughly the same values, Figure 10 displays the memory used for arbitrarily chosen key and message lengths.

\subsubsection{Power Consumption:}
The final metric to be analysed concerns power consumption values. Since the SNMAC was built targeting resource-constrained IoT devices, it is necessary to observe how many resources the algorithm uses over a period of running time. Similar to the memory experiment, the SNMAC cannot be successfully benchmarked against the HMAC construction for the same reasons mentioned above: it is difficult to isolate the resources used by the HMAC algorithm provided by an external library. Furthermore, the experimental results are not entirely accurate, as it is a challenging task to completely isolate and observe the execution of the algorithm inside the hardware of the device. To elaborate, the device will also execute other tasks, be it OS-related tasks or tasks issued by other running software. Thus, this experimental data only paints an approximate picture of the resources used by the SNMAC algorithm.

This power consumption experiment involves generating secret keys of randomly chosen lengths and hashing randomly chosen messages using the generated keys. This procedure is repeated throughout the course of 60 minutes, with no pauses. The execution of the algorithm is not parallelised in any way - this excludes the uses of concurrency, multiple threads or processes. Finally, data is collected using the external USB power monitor tool. The end result showed that constant runtime for 60 minutes uses $2.82 Wh$ (Watts / hour) which makes the prototyped solution attractive for low power environments.

\subsection{Limitations} 

To properly assess the quality and relevance of the experimental results, it is necessary to identify and discuss any limitations encountered throughout the performance analysis. Since the experiments were not conducted in an ideal environment, errors may occur, or details may be missed. 

First of all, it is vital to address the issues regarding benchmarking. While benchmarking the proposed SNMAC against the popular HMAC algorithm, this endeavour turned out to be more difficult than anticipated. The main difficulty factor originates from the source of the HMAC implementation, namely external libraries. External libraries were used because they provide ease of implementation, fast integration and, most of the time, are community tested and proven to work. However, there are some disadvantages of using external libraries: not being able to customize the source code or not being able to isolate the execution of a particular chunk of code. Furthermore, external libraries, especially the most popular ones, receive so much attention from the community that they end up being overly efficient in terms of execution time and resources used. To exemplify, the difference between the average running time of the SNMAC and HMAC algorithms can be easily justified by how many optimizations are involved in the implementation of the HMAC code.

Furthermore, the hardware and software choices only partially represent the resource-constrained IoT environment. While the Raspberry Pi 3 is considered an IoT device, it does possess much more computational power when compared to other, usually smaller, IoT devices. However, for practical reasons, it was deemed unfeasible to attempt using less accessible or user-friendly IoT devices, even though they would have been more representative.

Finally, it should be mentioned that the performance data was observed for the execution of a SNMAC prototype and, without proper optimizations and improvements, it is not entirely representative of the algorithm's potential. While this data paints a moderate picture of how the algorithm might behave in real-life scenarios, it cannot be successfully used to determine its feasibility.

\section{Security Analysis} \label{security_analysis}
In this section, we conclude by investigating the security of our SNMAC algorithm. We start with an experimental investigation of the avalanche effect, followed by a theoretical analysis of the properties inherited by the HMAC construction and the VSH compression function.

\subsection{Experimental Approach: Avalanche Effect}
We start with an experimental analysis of the avalanche effect of our SNMAC algorithm. The primary goal is secure input-to-output mapping with high randomization for improved security, minimizing traces available for attackers.

The avalanche effect in hashing means minor input changes lead to significant output changes. For example, changing one bit in an input should result in approximately half of the output bits changes, making the new hash very different from the original. To evaluate this property, we examined the Hamming distances between the outputs of the original inputs and the outputs of the modified ones, aiming for them to match those of random integers of the same length. We performed this experiment multiple times with randomized inputs, observing similarities to a normal probability density function.

\subsubsection{Experimental setup and results}

To run the experiments and collect data to observe the avalanche effect, the experimental setup used was identical to the one used in the performance analysis. Please refer to Section \ref{setup} for more details about the hardware and software environments.

In this experiment, the goal was to generate and modify random integers, hash them, and observe output changes. Various output bit lengths were considered. Before discussing the experiment, we need to understand the input-output relationship in the SNMAC algorithm. Section \ref{blueprint} reveals that the input is a pair $(n, m)$ (secret key and message), and the output is the tag, the hash for this pair. This means that the tag directly relates to the input $(n, m)$ pair. However, as described in Section \ref{hashing_function}, the output space depends only on the secret key $n$ and not the chosen $m$. In other words, the number of possible hashes depends solely on the key size, not the input message. To observe output changes with different bit lengths, we only need to adjust the key size without restricting the input message. In what follows, we describe the experimental setup in detail:

\begin{enumerate}
    \item Different secret key lengths are considered, determined by the number of primes used: 10, 20, 30, 40, 50, 60, 70, 80, 90. Each key generates one iteration.
    \item In each iteration, 1,000,000 random 256-bit input messages are generated, along with a modified version by flipping one random bit. This yields 1,000,000 pairs $(m, m')$.
    \item Each pair $(m, m')$ in an iteration is hashed, producing 1,000,000 pairs $(h, h')$.
    \item Hamming distances between "$h$" and "$h'$" are computed and recorded, resulting in 1,000,000 distances per iteration. The average length of "$h$" and "$h'$" is also recorded as $\Delta$.
    \item In each iteration, a density histogram of the Hamming distances is plotted against a normal probability density function using the recorded mean and variance.
\end{enumerate}

\begin{figure}[t]
    \centering
    \includegraphics[width=\linewidth]{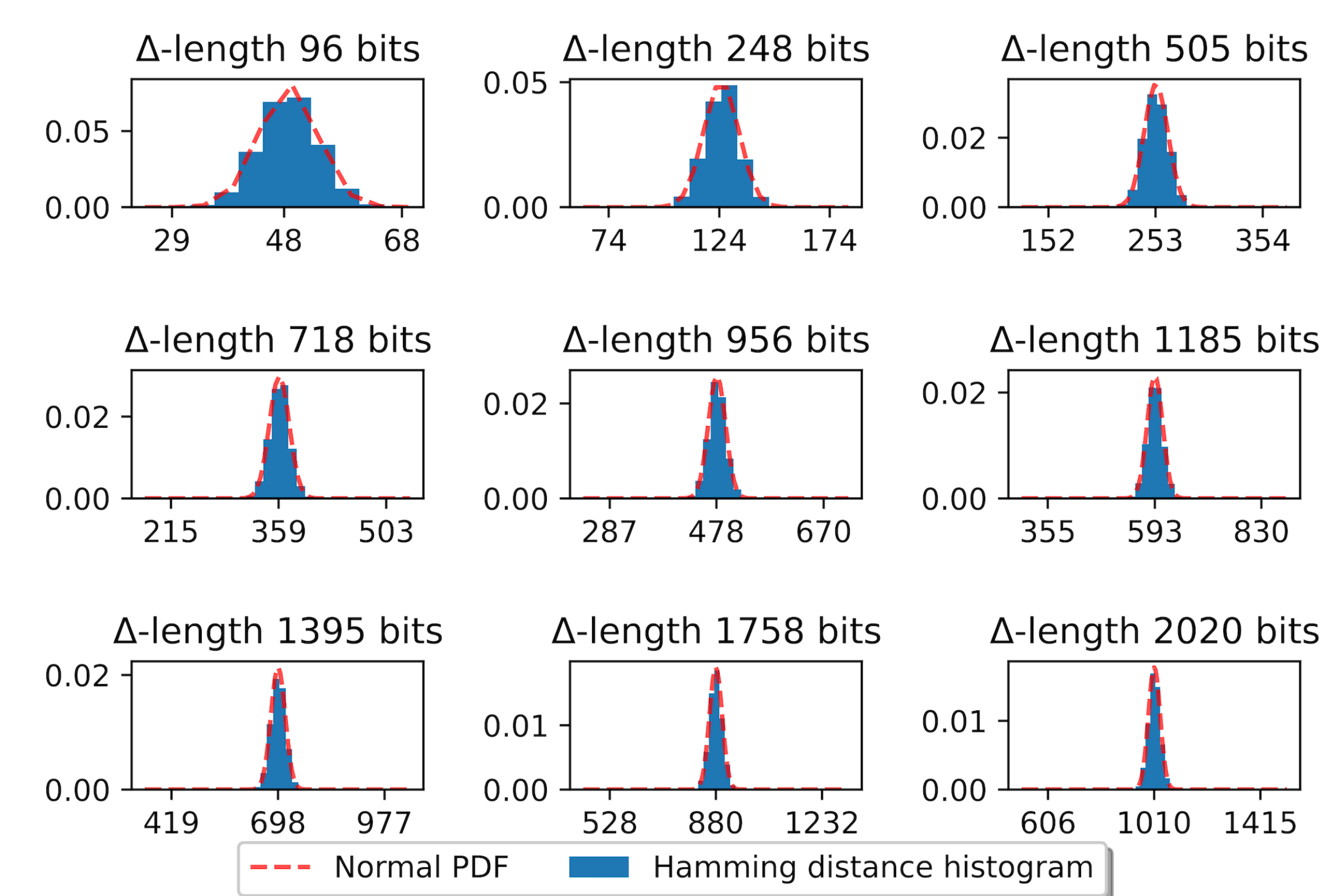}
    \caption{Average Observed Hamming Distances of Hashes of various lengths}
    \label{fig:hamming}
\end{figure}

The experiment results are shown in Figure \ref{fig:hamming}. Average hash length increases with the number of primes in the secret key, as expected. Hamming distance data and normal probability distribution plots are observed for each experiment, with the histogram data fitting the normal distribution. Importantly, both plots have a peak at roughly half the average hash length, which is the desired outcome. This demonstrates significant output changes without revealing meaningful statistical information about the input messages.

\subsection{Theoretical discussions}

The second part of the security analysis concerns the theoretical point of view. While the experimental approach showed promising results, it is important to understand that the successful property of the avalanche effect does not merely determine the security of a hashing construct. To completely address the security implications, it is required to delve into the security requirements of a MAC (and HMAC) and other theoretical hash function properties: pre-image resistance, second pre-image resistance and collision resistance. 

\subsubsection{MAC and HMAC security requirements}

A MAC algorithm is considered secure if it can resist an existential forgery under a chosen-message attack (EUF-CMA). Existential forgery implies the ability of an attacker to create at least one message-hash pair where the message has never been legitimately hashed. The attacker can freely select any message as long as the message-hash pair is valid. The "chosen-message" part can be exemplified in the following security game: an attacker has access to an oracle, which can generate a secret key (unknown to the attacker) and can also compute a MAC for any message chosen by the attacker. The attacker is considered unable to win the game if they cannot generate any valid hash for a chosen message other than the ones used when accessing the oracle without using unreasonable amounts of computational resources. However, HMAC aims to satisfy the EUF-CMA security requirements by design:

\begin{itemize}
    \item EUF-CMA implies that the security key is hard to recover. Besides the obvious key leaking scenarios, the HMAC construction makes extracting the key from the generated hash difficult. Recall that the key is both concatenated with two distinct values, as well as hashed twice during the generation procedure. 

    \item EUF-CMA implies that it is challenging to perform selective forgery. Thus, it must be difficult to forge a tag on a specific message chosen by the attacker. Since the generated hash directly depends on the input message and the secret key, this ties to the previous requirement: the difficulty of retrieving the key makes selective forgery difficult.

    \item EUF-CMA implies that it is difficult to forge a new hash for an already existing message-hash pair. Since each hash generated is ideally unique, this implies that forgery must be made for a new message.
\end{itemize}

While HMAC makes the security analysis easier by meeting critical MAC security requirements, it comes with its own restrictions. This is very useful since the SNMAC algorithm is a variant of the HMAC algorithm, with a substituted hashing function. The cryptographic strength of HMAC heavily relies on the bit length of the secret key (and, in turn, on the length of the generated hashes and the total output space) and on the security properties of its hashing function \cite{DBLP:journals/rfc/rfc2104}. The first criterion is outside the scope of this analysis because it is mostly the communicating parties' responsibility. As with most other keyed security schemes, it is advised to increase the key length as much as possible to increase the complexity of brute-force attacks or other derivatives. The second criterion is of more interest to the analysis and will be expanded upon in the following subsection.             

\subsubsection{Hashing function security requirements}

The security of HMAC directly depends on the secret key length and the security properties of the underlying hashing function \cite{DBLP:journals/rfc/rfc2104}. Since SNMAC is a variant of HMAC, the security principles apply in the same way. Thus, our analysis looks at the security levels of the VSH used inside the SNMAC construct. Namely, we focus on the three core properties of a hash function:

\begin{itemize}
    \item \textit{Pre-image resistance}: given a hash $h$, it should be difficult to find a message $m$ such that $h = H(m)$.

    \item \textit{Second pre-image resistance}: given an input message $m$, it should be difficult to find a message $m'$ such that $H(m) = H(m')$.

    \item \textit{Collision resistance}: it should be difficult to find two messages $m$ and $m'$ such that $H(m) = H(m')$.
\end{itemize}

The original authors of the VSH algorithm have already discussed and proven the collision resistance of the VSH under the VSSR assumption (recall Section \ref{hashing_function}) \cite{DBLP:conf/eurocrypt/ContiniLS06}. This collision resistance implies that VSH is also second pre-image resistant. However, the original authors make no note of the pre-image resistance property. Instead, a later study shows that VSH is clearly not pre-image resistant due to its multiplicative property \cite{DBLP:conf/indocrypt/Saarinen06}. The authors showed that three bit strings of equal length $x,y,z$ can be chosen such that $z$ consists of only 0 bits and $x \land y = z$. In such a scenario, $H(z)H(x \lor y) = H(x)H(y) (\mod n)$. This illustrates that cracking the hash can be done effectively in "square-rooted" time, i.e. $n$-bit hashes are as easy to break as $n/2$-bit hashes. A more efficient attack has been discovered by other authors as well \cite{DBLP:conf/IEEEares/HalunenRR09}. This, unfortunately, also implies that SNMAC has, to an extent, a multiplicative property and can suffer from faster and more efficient brute-forcing attacks.

\subsubsection{Theoretical results}

The results of this analysis are two-fold. On the bright side, the SNMAC borrows favourable properties from HMAC and achieves most of the security requirements, including being resistant against EUF-CMA. Conversely, the proposed algorithm suffers from its underlying VSH function due to its multiplicative properties, which undermines pre-image resistance.

However, is the lack of pre-image resistance a deal breaker for the security of SNMAC? Several arguments can be made in SNMAC's favour:

\begin{enumerate}
    \item While VSH is not pre-image resistant due to its multiplicative nature, SNMAC introduces more computational complexity due to the execution of two hashing passes and the inner concatenations. It is uncertain if this increase in complexity makes the attacking process more difficult; thus, it is a question for further studies.
    
    \item A recent study showed that HMAC is a pseudo-random function if the hashing compression function, assuming a Merkle-Damaard construction, is also a pseudo-random function \cite{DBLP:journals/joc/Bellare15}. Since VSH follows the Merkle-Damg\r{a}rd construction, the VSH's compression function can be studied in future research to determine if it is a pseudo-random function or, if possible, potentially transformable into one. 

    \item Several variants of VSH could be of interest for future research since they might be able to solve the multiplicative properties. The original authors proposed several of them, including a cubing algorithm and a discrete logarithm-based compression function \cite{DBLP:conf/eurocrypt/ContiniLS06}. Other works analyse "smoother" versions \cite{DBLP:conf/acisp/Sarinay11} or elliptic-curve implementations \cite{DBLP:conf/vietcrypt/LenstraPS06}. These works show that improvements can be made, which may lead to a more secure hashing function.
\end{enumerate}

\section{Conclusion}
\label{conclusion}
The IoT infrastructure is in constant demand of proper security tools, including authentication mechanisms, while, at the same time, being heavily constrained by available hardware. Thus, efficient security algorithms must be designed with the particular needs and restrictions of IoT devices. Although academia has discussed this topic before, one detail has yet to be thoroughly explored: smooth numbers used for optimization. 

In this study, a Smooth Number Message Authentication Code (SNMAC) algorithm is designed and proposed as a valuable variant of the classical Hash-Based Message Authentication Code (HMAC). This alternative solution is prototyped, tested in terms of performance and analysed in terms of security. The results shown are promising, encouraging further studies into the use of smooth numbers for cryptographic tasks. However, there are both security and performance concerns which are yet to be addressed before certifying this alternative solution for practical use in real scenarios.

\bibliographystyle{abbrv}
\bibliography{bibliography}

\begin{thebibliography}{10}

\bibitem{DBLP:conf/tgc/AmarilliHBMNR12}
A.~Amarilli, F.~B. Hamouda, F.~Bourse, R.~Morisset, D.~Naccache, and P.~Rauzy.
\newblock From rational number reconstruction to set reconciliation and file
  synchronization.
\newblock In C.~Palamidessi and M.~D. Ryan, editors, {\em Trustworthy Global
  Computing - 7th International Symposium, {TGC} 2012, Newcastle upon Tyne, UK,
  September 7-8, 2012, Revised Selected Papers}, volume 8191 of {\em Lecture
  Notes in Computer Science}, pages 1--18. Springer, 2012.

\bibitem{aragona2016real}
R.~Aragona, F.~Gozzini, and M.~Sala.
\newblock A real life project in cryptography: assessment of rsa keys.
\newblock In {\em Physical and Data-Link Security Techniques for Future
  Communication Systems}, pages 197--203. Springer, 2016.

\bibitem{DBLP:journals/joc/Bellare15}
M.~Bellare.
\newblock New proofs for {NMAC} and {HMAC:} security without collision
  resistance.
\newblock {\em J. Cryptol.}, 28(4):844--878, 2015.

\bibitem{briggs1998introduction}
M.~E. Briggs.
\newblock {\em An introduction to the general number field sieve}.
\newblock PhD thesis, Virginia Tech, 1998.

\bibitem{DBLP:conf/green/CastellonRKDB22}
C.~Castellon, S.~Roy, O.~P. Kreidl, A.~Dutta, and L.~B{\"{o}}l{\"{o}}ni.
\newblock Towards an energy-efficient hash-based message authentication code
  {(HMAC)}.
\newblock In {\em 13th {IEEE} International Green and Sustainable Computing
  Conference, {IGSC} 2022, Pittsburgh, PA, USA, October 24-25, 2022}, pages
  1--7. {IEEE}, 2022.

\bibitem{DBLP:conf/eurocrypt/ContiniLS06}
S.~Contini, A.~K. Lenstra, and R.~Steinfeld.
\newblock Vsh, an efficient and provable collision-resistant hash function.
\newblock In S.~Vaudenay, editor, {\em Advances in Cryptology - {EUROCRYPT}
  2006, 25th Annual International Conference on the Theory and Applications of
  Cryptographic Techniques, St. Petersburg, Russia, May 28 - June 1, 2006,
  Proceedings}, volume 4004 of {\em Lecture Notes in Computer Science}, pages
  165--182. Springer, 2006.

\bibitem{cooley1965algorithm}
J.~W. Cooley and J.~W. Tukey.
\newblock An algorithm for the machine calculation of complex fourier series.
\newblock {\em Mathematics of computation}, 19(90):297--301, 1965.

\bibitem{DBLP:journals/tc/DimitrovVA22}
V.~S. Dimitrov, L.~Vigneri, and V.~Attias.
\newblock Fast generation of {RSA} keys using smooth integers.
\newblock {\em {IEEE} Trans. Computers}, 71(7):1575--1585, 2022.

\bibitem{DBLP:conf/nspw/Durmuth13}
M.~D{\"{u}}rmuth.
\newblock Useful password hashing: how to waste computing cycles with style.
\newblock In M.~E. Zurko, K.~Beznosov, T.~Whalen, and T.~Longstaff, editors,
  {\em New Security Paradigms Workshop, {NSPW} '13, Banff, AB, Canada,
  September 9-12, 2013}, pages 31--40. {ACM}, 2013.

\bibitem{DBLP:journals/fi/ElhajjMF23}
M.~El{-}hajj, H.~Mousawi, and A.~Fadlallah.
\newblock Analysis of lightweight cryptographic algorithms on iot hardware
  platform.
\newblock {\em Future Internet}, 15(2):54, 2023.

\bibitem{granville2008smooth}
A.~Granville.
\newblock Smooth numbers: computational number theory and beyond.
\newblock {\em Algorithmic number theory: lattices, number fields, curves and
  cryptography}, 44:267--323, 2008.

\bibitem{DBLP:conf/IEEEares/HalunenRR09}
K.~Halunen, P.~Rikula, and J.~R{\"{o}}ning.
\newblock Finding preimages of multiple passwords secured with {VSH}.
\newblock In {\em Proceedings of the The Forth International Conference on
  Availability, Reliability and Security, {ARES} 2009, March 16-19, 2009,
  Fukuoka, Japan}, pages 499--503. {IEEE} Computer Society, 2009.

\bibitem{DBLP:journals/sj/HammiFKZB20}
B.~Hammi, A.~Fayad, R.~Khatoun, S.~Zeadally, and Y.~Begriche.
\newblock A lightweight ecc-based authentication scheme for internet of things
  (iot).
\newblock {\em {IEEE} Syst. J.}, 14(3):3440--3450, 2020.

\bibitem{karthi2019enhanced}
G.~Karthi and M.~Ezhilarasan.
\newblock Enhanced vsdl hash algorithm for data integrity and protection.
\newblock In {\em Data Management, Analytics and Innovation: Proceedings of
  ICDMAI 2018, Volume 2}, pages 527--539. Springer, 2019.

\bibitem{DBLP:journals/rfc/rfc2104}
H.~Krawczyk, M.~Bellare, and R.~Canetti.
\newblock {HMAC:} keyed-hashing for message authentication.
\newblock {\em {RFC}}, 2104:1--11, 1997.

\bibitem{DBLP:conf/vietcrypt/LenstraPS06}
A.~K. Lenstra, D.~Page, and M.~Stam.
\newblock Discrete logarithm variants of {VSH}.
\newblock In P.~Q. Nguyen, editor, {\em Progressin Cryptology - {VIETCRYPT}
  2006, First International Conference on Cryptology in Vietnam, Hanoi,
  Vietnam, September 25-28, 2006, Revised Selected Papers}, volume 4341 of {\em
  Lecture Notes in Computer Science}, pages 229--242. Springer, 2006.

\bibitem{DBLP:journals/ijnsec/PadmavathyB15}
R.~Padmavathy and C.~Bhagvati.
\newblock A new method for computing {DLP} based on extending smooth numbers to
  finite field for ephemeral key recovery.
\newblock {\em Int. J. Netw. Secur.}, 17(3):243--254, 2015.

\bibitem{DBLP:conf/indocrypt/Saarinen06}
M.~O. Saarinen.
\newblock Security of {VSH} in the real world.
\newblock In R.~Barua and T.~Lange, editors, {\em Progress in Cryptology -
  {INDOCRYPT} 2006, 7th International Conference on Cryptology in India,
  Kolkata, India, December 11-13, 2006, Proceedings}, volume 4329 of {\em
  Lecture Notes in Computer Science}, pages 95--103. Springer, 2006.

\bibitem{DBLP:conf/acisp/Sarinay11}
J.~Sarinay.
\newblock Faster and smoother - {VSH} revisited.
\newblock In U.~Parampalli and P.~Hawkes, editors, {\em Information Security
  and Privacy - 16th Australasian Conference, {ACISP} 2011, Melbourne,
  Australia, July 11-13, 2011. Proceedings}, volume 6812 of {\em Lecture Notes
  in Computer Science}, pages 142--156. Springer, 2011.

\bibitem{DBLP:conf/giots/Suarez-AlbelaFF18}
M.~Su{\'{a}}rez{-}Albela, T.~M. Fern{\'{a}}ndez{-}Caram{\'{e}}s,
  P.~Fraga{-}Lamas, and L.~Castedo.
\newblock A practical performance comparison of {ECC} and {RSA} for
  resource-constrained iot devices.
\newblock In {\em 2018 Global Internet of Things Summit, GIoTS 2018, Bilbao,
  Spain, June 4-7, 2018}, pages 1--6. {IEEE}, 2018.

\bibitem{vailshery_2016}
L.~S. Vailshery.
\newblock Number of iot devices 2015-2025, Nov 2016.

\end{thebibliography}

\end{document}